\documentclass[conference]{IEEEtran}

\usepackage{amssymb}
\usepackage[cmex10]{amsmath}
\usepackage{stfloats}
\usepackage{graphicx}
\usepackage{subfigure}
\usepackage{tabularx}
\usepackage{epsfig,epsf,color,balance,cite}
\usepackage{verbatim}
\usepackage{url}
\usepackage{bm}

\newtheorem{theorem}{Theorem}
\newtheorem{lemma}{Lemma}

\usepackage{algorithm,algorithmic}

\begin{document}

\title{Resource Allocation in Full-Duplex Mobile-Edge Computing Systems with NOMA and \\ Energy Harvesting}

\author{
\IEEEauthorblockN{Zhaohui Yang,
Jiancao Hou
                  and Mohammad Shikh-Bahaei
                  }
\IEEEauthorblockA{Centre for Telecommunications Research, Department of Informatics, King's college London, UK}
\IEEEauthorblockA{E-mail: \{yang.zhaohui, jiancao.hou, m.sbahaei\}@kcl.ac.uk} }

\maketitle

\begin{abstract}
This paper considers a full-duplex (FD) mobile-edge computing (MEC) system with non-orthogonal multiple access (NOMA) and energy harvesting (EH), where one group of users simultaneously offload task data to the base station (BS) via NOMA and the BS simultaneously receive data and broadcast energy to other group of users with FD. We aim at minimizing the total energy consumption of the system via power control, time scheduling and computation capacity allocation. To solve this nonconvex problem, we first transform it into an equivalent problem with less variables. The equivalent problem is shown to be convex in each vector with the other two vectors fixed, which allows us to design an iterative algorithm with low complexity. Simulation results show that the proposed algorithm achieves better performance than the conventional methods.
\end{abstract}

\IEEEpeerreviewmaketitle

\section{Introduction}

Mobile-edge computing (MEC) has been deemed as a promising technology for future communications due to that it can improve the computation capacity of users in applications, such as, augmented reality (AR) \cite{8016573}.
With MEC, users can offload the tasks to the MEC servers that locate at the edge of the system.
Since the MEC servers can be deployed near to the users, system with MEC can provide users with low energy consumption and low latency\cite{7906521}.

The basic idea of MEC is to utilize the powerful computing facilities within the radio access system, such as the MEC server integrated into the base station (BS).
Users can benefit from offloading the computationally intensive tasks to the MEC server.
There are two operation modes for MEC, i.e., partial and binary computation offloading.
In partial computation offloading, the computation tasks can be divided into two parts, where one part is locally executed and the other part is offloaded to the MEC servers \cite{8006982,8254208,7762913,7929399,8240666,8168252,7842016}.
In binary computation offloading, the computation tasks are either locally executed or offloaed to the MEC servers \cite{6574874}.

In wireless systems, the system performance is always constrained due to limited battery capacity of users. 
To prolong the system lifetime, many contributions \cite{zhou2014wireless,YuanWireless,6678102,Li2017Sum,Yang2018Optimal} investigate energy harvesting (EH), where users can harvest energy in a wireless way from the dedicated energy transmitter.
Combining EH with MEC is a promising technique to provide sustainable computation experience for users. 
Due to the fact that EH generally occupies non-negligible bandwidth, full-duplex (FD) \cite{towhidlou2018improved,naslcheraghi2017fd,nehra2010cross,towhidlou2016cooperative} can be applied to improve the spectral efficiency by means of simultaneous energy transmission and computation task offloading in the same bandwidth \cite{7934379}.
Integrating EH and FD technologies into MEC, the max-min energy efficiency optimization problem was investigated for a FD-MEC system with EH in \cite{8254208}.

Recently, non-orthogonal multiple access (NOMA) has been recognized as a potentional technology for the next generation mobile communication systems to tackle the explosive growth of data traffic \cite{saito2013non,Zhiguo2017Survey,7263349,Yang2017Fair,Yang2017On,Yang2017Energy,8319496}.
Due to superposition coding at the transmitter and successive interference cancelation (SIC) at the receiver, NOMA can achieve higher spectral efficiency than conventional orthogonal multiple access (OMA), such as time division multiple access (TDMA) and orthogonal frequency division multiple access (OFDMA).
Many previous contributions \cite{7906521,8006982,8254208,7762913,7929399,8240666,8168252,7842016} only considered OMA.
Motivated by the benefits of NOMA over OMA, a NOMA-based MEC system was investigated in \cite{8269088}, where users simultaneously offload their computation tasks to the BS and the BS uses SIC for information decoding.
Besides, both NOMA uplink and downlink transmissions were applied to MEC \cite{zhi2018ImNOMAMEC}, where analytical results were developed to show that  the latency and energy consumption can be reduced by applying NOMA-based MEC offloading.
The benefits of NOMA and EH were investigated in \cite{7946258,Yang2018EEIoT,8116367}.
However, the above NOMA-based MEC systems \cite{8269088,zhi2018ImNOMAMEC} did not consider EH even though EH can further prolong the lifetime of the system.
To our best knowledge, FD-MEC systems with NOMA and EH have not been investigated in the literature.

In this paper, we investigate the resource allocation in a FD-MEC system with NOMA and EH, where users simultaneously offload computation tasks to the BS through NOMA and the BS simultaneously broadcast energy and receive computation tasks via FD.
The main contributions of this paper are summarized as follows:
\begin{enumerate}
  \item  The total energy consumption of the system is formulated for a FD-MEC system with NOMA and EH via power control, time scheduling and offloading data allocation.
  \item  By using the recursion method, the uplink transmission power of each user can be presented as a function with scheduled time, transmission power of the BS and offloading data. Based on this finding, the original problem can be equivalent to a problem with less variables.
  \item The equivalent problem is proved to be convex in power vector or time vector or offloading data vector with the other two vectors fixed. Owing to this characteristic, an iterative algorithm is accordingly proposed with low complexity.
\end{enumerate}

The rest of the paper is organized as follows.
In Section $\text{\uppercase\expandafter{\romannumeral2}}$, we introduce the system model and formulate the total energy minimization problem.
Section $\text{\uppercase\expandafter{\romannumeral 3}}$ provides the optimal conditions and an iterative algorithm.
Some numerical results are shown in Section $\text{\uppercase\expandafter{\romannumeral 4}}$
and conclusions are finally drawn in Section $\text{\uppercase\expandafter{\romannumeral5}}$.

\section{System Model and Problem Formulation}
\subsection{System Model}
\begin{figure}
\centering
\includegraphics[width=3.0in]{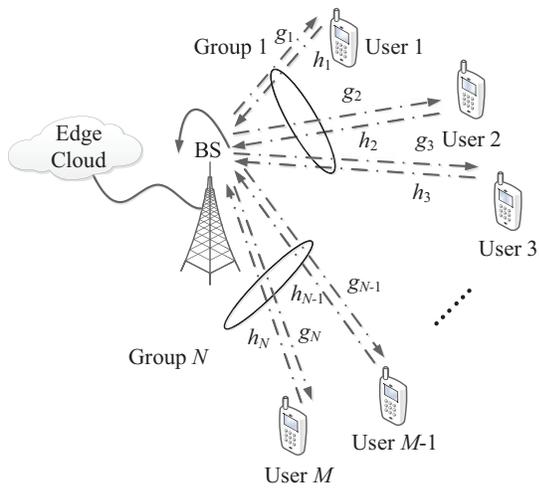}
\vspace{-1.5em}
\caption{Multi-user MEC system.}\label{sys1}
\vspace{-1.5em}
\end{figure}

Consider a multi-user FD MEC system with $M$ users and one BS that is the gateway of an edge cloud, as shown in Fig.~\ref{sys1}.
To perform NOMA, all users are classified into $N$ small groups.
Denote the sets of users and groups by $\mathcal M=\{1, \cdots, M\}$ and $\mathcal N=\{1, \cdots, N\}$, respectively.
The set of users in group $i$ is denoted by ${\cal{J}}_i=\{J_{i-1}+1, \cdots, J_i\}$, where $J_0=0$, $J_N=M$, $J_i=\sum_{l=1}^i|{\cal{J}}_l|$, and $|\cdot|$ is the cardinality of a set.
Obviously, we have $\bigcup_{i\in\mathcal N} \mathcal J_i=\mathcal M$.

\begin{figure}
\centering
\includegraphics[width=3.0in]{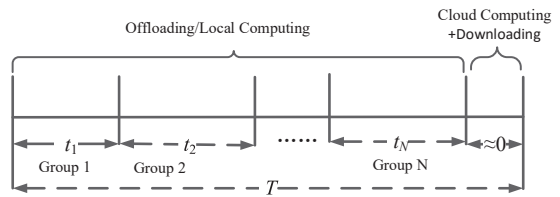}
\vspace{-1.5em}
\caption{Transmission period.}\label{sys2}
\vspace{-1.5em}
\end{figure}

The time slot with duration $T$ is divided into $N$ phases, as shown in Fig.~\ref{sys2}.
Note that the computation latency at the BS and downloading time of computation results are low and negligible \cite{7842016}.
In the $i$-th phase with time $t_i$, the users in group $i$ simultaneously communicate with the BS by using NOMA.
For user $j$, it is required to transmit $R_j$-bits input data within the time slot.
To save energy and meet the latency constraint, user $j$ offload $d_j$ bits out of $R_j$ bits to the BS.
Besides, the BS has fixed energy supply, while users do not have stable energy supply and need to harvest energy from the BS.
During the whole transmission phase, the BS keeps transforming energy to users.

Let $h_j$ denote the uplink channel gain between user $j$ and the BS.
Without loss of generality, the uplink channels between users in group $i$ and the BS are sorted as $h_{J_{i-1}+1} \geq \cdots \geq h_{ J_i}$.
With NOMA and FD technologies, the uplink achievable rate for user $j\in\mathcal J_i$ is
\begin{equation}\label{sys1eq1}
r_{ij}=B \log_2\left(
1+\frac{h_{j} p_j}
{ \sum_{l=j+1}^{J_i} h_{l} p_l +\sigma^2+\gamma q_i}
\right),
\end{equation}
where $B$ is the bandwidth of the system, $p_j$ is the transmission power of user $j$,
$\sigma^2$ represents the noise power,
$\gamma$ is the self-interference coefficient at the BS,
and $q_i$ is the broadcasting power of the BS in the $i$-th phase.
In equation (\ref{sys1eq1}), $\gamma q_i$ represents the residual self-interference at the BS due to the finite receiver dynamic range and imperfect channel estimation.
To successfully offload $d_j$ bits to the BS for user $j$, we have
\begin{equation}\label{sys1eq2}
r_{ij } t_{i} \geq d_j, \quad \forall j\in\mathcal J_i.
\end{equation}

Considering uplink transmission energy
at the user side,
the energy consumption for offloading at user $j\in\mathcal J_i$ is
\begin{equation}\label{sys1eq5}
E_{ij}^{\text{Off}}=p_j t_i. 
\end{equation}
Since only $d_j$ bits are offloaded to the BS, the remaining $R_j-d_j$ bits are needed to be computed locally at user $j$.
Based on the local computing model in  \cite{7842016}, the total energy consumption for local computation at user $j\in\mathcal J_i$ is given by
\begin{equation}\label{sys1eq6}
E_{ij}^{\text{Loc}} = (R_j-d_j) C_jP_j,
\end{equation}
where $C_j$ is the number of CPU cycles required for computing 1-bit input data at user $j$,
and $P_j$ stands for the energy consumption per cycle for local computing at this user.

Due to the fact that the BS broadcasts energy to users all the time, the energy harvested by user $j$ in group $i$ is given by \cite{zhou2014wireless,YuanWireless,6678102}
\begin{equation}\label{sys1eq23}
E_{ij}^{\text{H}}=\zeta_j g_j \sum_{k\in \mathcal N\setminus\{ i\} } q_i t_i, \quad \forall j\in\mathcal J_i,
\end{equation}
where $\zeta_j$ is the energy efficiency of the EH process for user $j$, and $g_j$ is the channel gain between the BS and user $j$.
According to the energy causality constraint in EH systems, the harvested energy should no less than the consumed energy for user $j$, i.e.,
\begin{equation}\label{sys1eq6_1}
E_{ij}^{\text{H}}\geq E_{ij}^{\text{Off}} +E_{ij}^{\text{Loc}}.
\end{equation}

Denote $F_j$ as the  computation capacity of user $j$, which is measured by the number of CPU cycles per second.
To meet the computation latency, we have
\begin{equation}\label{sys1eq7}
C_j (R_j-d_j) \leq F_j T.
\end{equation}

For the BS, the total energy consumption includes both the broadcasting and the the energy consumption for computation.
As a result, the total energy consumption of the BS is
\begin{equation}\label{sys1eq7_2}
E^{\text{BS}} = \sum_{i=1}^N q_i t_i +P_0 \sum_{j=1}^{M} C_j d_j,
\end{equation}
where $P_0$ is the energy consumption per cycle at the BS.
The first term in the left-hand side of (\ref{sys1eq7_2}) is the broadcasting energy, while the second term in the left-hand side of (\ref{sys1eq7_2}) represents the computation energy.

Based on (\ref{sys1eq5}), (\ref{sys1eq6}), (\ref{sys1eq23}) and  (\ref{sys1eq7_2}), the total energy consumption of the system can be given by
\begin{equation}\label{sys1eq3_1}
E^{\text{Total}} = E^{\text{BS}}+ \sum_{i=1}^N\sum_{j=J_{i-1}+1}^{J_i} (E_{ij}^{\text{Off}}+E_{ij}^{\text{Loc}}-E_{ij}^{\text{H}}).
\end{equation}

For edge cloud, it is assumed that the edge cloud has finite computation capacity, denoted as $F$, measured
as the maximum CPU cycles allowed for computing the sum offloaded data in each slot, i.e.,
\begin{equation}\label{sys1eq8}
\sum_{j=1}^{M} C_j d_j  \leq F, 
\end{equation}
which ensures low computing time at the edge cloud.

\subsection{Problem Formulation}
Now it is ready to investigate the total energy minimization problem.
Mathematically, it is formulated as
\begin{subequations}\label{sysmin1}
\begin{align}
 \mathop{\min}_{\pmb{p}, \pmb q, \pmb t, \pmb d}\:\;\quad
&  E^{\text{Total}}
\\
\textrm{s.t.}\qquad
&r_{ij } t_{i} \geq d_j, \quad\forall i\in \mathcal N, j \in \mathcal J_i\\
&E_{ij}^{\text{H}}\geq E_{ij}^{\text{Off}} +E_{ij}^{\text{Loc}}, \quad\forall i\in \mathcal N, j \in \mathcal J_i\\
& C_j (R_j-d_j) \leq F_j T, \quad \forall i \in \mathcal N, j\in \mathcal J_i\\
& \sum_{j=1}^{M} C_j d_j  \leq F\\
&\sum_{i=1}^{N} t_i \leq T \\
&0\leq p_j \leq P_j, q_i \leq Q, t_i \geq 0, \quad\forall i\in \mathcal N, j \in \mathcal J_i,
\end{align}
\end{subequations}
where
$\pmb p=[p_1, \cdots, p_M]^T$,
$\pmb q=[q_1, \cdots, q_N]^T$,
$\pmb t=[t_{1}, \cdots,$ $ t_{N}]^T$,
$\pmb d=[d_{1}, \cdots,  d_{M}]^T$,
$P$ is the maximal transmission power of each user and $Q$ is the maximal transmission power of the BS.
The objective function (\ref{sysmin1}a) is the total energy consumption of the system including transmission and computation energy.
Constraints (\ref{sysmin1}b) represent the the minimal transmitted data constraints for uplink.
The consumed energy of each user should not exceed its harvested energy, as stated in constraints (\ref{sysmin1}c).
The computation delay constraints for users to compute tasks locally are given in (\ref{sysmin1}d), while (\ref{sysmin1}e) ensures the low computing time at the BS.
Constraint (\ref{sysmin1}f) is the time division constraint.
The maximal transmission power limits for the BS and users are given in (\ref{sysmin1}g).

\section{Algorithm Design}
Due to nonconvex objective function ((\ref{sysmin1}a) and nonconvex constraints (\ref{sysmin1}b)-(\ref{sysmin1}c), total energy minimization problem (\ref{sysmin1}) is nonconvex. To solve this nonconvex problem, we first obtain the optimal conditions and then accordingly propose an iterative algorithm with low complexity.
\subsection{Optimal Conditions}
By analyzing problem (\ref{sysmin1}), we have the following lemma.
\begin{lemma}
The optimal ($\pmb p^*, \pmb q^*, \pmb t^*, \pmb d^*$) of problem (\ref{sysmin1}) satisfies the following conditions
\begin{eqnarray}
r_{ij } t_{i} = d_j, \quad\forall i\in \mathcal N, j \in \mathcal J_i \label{alg2eq1}.
\end{eqnarray}
\end{lemma}

Lemma 1 can be proved by using the contradictory method.
Obviously, we can show that  $r_{ij } t_{i} = d_j$ for the optimal solution, as otherwise (\ref{sysmin1}a) can be further improved by decreasing $p_{j}$ with all constraints satisfied, contradicting that the solution is optimal.
Lemma 1 states that transmitting with minimal number of data bits is optimal. This is intuitive since transmitting with less resource is always energy saving.

\subsection{Joint Power Control, Time Scheduling and Computation Capacity Allocation}
To solve nonconvex problem (\ref{sysmin1}), we first have the following theorem.

\begin{theorem}
Problem (\ref{sysmin1}) is equivalent to the following problem:
\begin{subequations}\label{alg2min2}
\begin{align}
 \mathop{\min}_{ \pmb q, \pmb t, \pmb d}\:\;
&   \!\sum_{i=1}^N \!q_i t_i \!+\!P_0 \!\sum_{j=1}^{M} \!C_j d_j
\!+\!\sum_{i=1}^N\!\sum_{j=J_{i-1}+1}^{J_i} \!t_i f_{ij}(t_i,q_i,\pmb d)
\nonumber\\&
+\!\sum_{i=1}^N\!\sum_{j=J_{i-1}+1}^{J_i} \!\left(\!
(R_j-d_j) C_jP_j-
\zeta_j g_j \sum_{k\in \mathcal N\setminus\{ i\} } q_i t_i
\!\right) \!
\\
\textrm{s.t.}\quad
&\zeta_j g_j \sum_{k\in \mathcal N\setminus\{ i\} } q_i t_i\geq  t_i f_{ij}(t_i,q_i,\pmb d) +(R_j-d_j) C_jP_j,
\nonumber \\
&\qquad\qquad\qquad\qquad\quad\forall i\in \mathcal N, j \in \mathcal J_i\\
& C_j (R_j-d_j) \leq F_j T, \quad \forall i \in \mathcal N, j\in \mathcal J_i\\
& \sum_{j=1}^{M} C_j d_j  \leq F  \\
&\sum_{i=1}^{N} t_i \leq T \\
&f_{ij}(t_i, q_i, \pmb d ) \leq P_j, \quad\forall i\in \mathcal N, j \in \mathcal J_i,
\\
&q_i \leq Q, t_i \geq 0, \quad\forall i\in \mathcal N,
\end{align}
\end{subequations}
where
\begin{eqnarray}\label{alg2eq2}
&&\!\!\!\!\!\!\!\!\!\!\!\!\!\!\!\!\!\! f_{ij}(t_i,q_i,\pmb d)
\triangleq\frac{(\sigma^2+\gamma q_i)}
{h_{j}}
\left(2^{\frac{d_{j}}{B t_i}} -1\right)\nonumber\\
&&\!\!\!\!\!\!\!\!\!\!\!\!\!\!\!\!\!\!\quad+
\sum_{l=j+1}^{J_i}
\frac{(\sigma^2+\gamma q_i)}
{h_{j} }
\!\left(\!2^{\frac{d_{l}}{B t_i}} \!-\!1\!\right)\! \left(\!{2^{\frac{d_{j}}{B t_i}}}\!-\!1\!\right)\!{2^{\frac{\sum_{s=j+1}^{l-1}d_{s}}{B t_i}}}
.
\end{eqnarray}
\end{theorem}

\itshape \textbf{Proof:}  \upshape
Please refer to Appendix A.
\hfill $\Box$

In (\ref{alg2eq2}), $f_{ij}(t_i, q_i, \pmb d)$ is the transmission power of user $j$ in group $i$, which is shown to be a function of scheduled time, transmission power of the BS and offloading data. 
According to Theorem 1, problem (\ref{sysmin1}) can be simplified by solving an equivalent problem (\ref{alg2min2}) with less variables.
By analysing problem (\ref{alg2min2}), we can obtain the following theorem.
\begin{theorem}
Problem (\ref{alg2min2}) is convex in each variable with the other two variables fixed, i.e., problem (\ref{alg2min2}) is convex in $\pmb q$ with fixed $(\pmb t, \pmb d)$, $\pmb t$ with fixed $(\pmb q, \pmb d)$, and $\pmb d$ with fixed $(\pmb q, \pmb t)$.
\end{theorem}

\itshape \textbf{Proof:}  \upshape
Please refer to Appendix B.
\hfill $\Box$

Based on theorem 1, we can easily optimize each variable with the other two variables fixed through solving a correspondingly convex problem, which can be solved by using the popular interior method \cite[Page~561]{boyd2004convex}.
Owing  to this characteristic, we can propose
 an iterative algorithm to effectively solve problem (\ref{alg2min2}) in Algorithm 1, i.e., iterative power control, time scheduling and offloading data allocation algorithm.

\begin{algorithm}[h]
\caption{\!\!: Iterative Power Control, Time Scheduling and Offloading Data Allocation Algorithm}
\begin{algorithmic}[1]
\STATE Set the initial solution $( \pmb q^{(0)}, \pmb t^{(0)}, \pmb d^{(0)})$, and iteration number $n=1$.
\REPEAT
\STATE With fixed $( \pmb t^{(n-1)}, \pmb d^{(n-1)})$, obtain the optimal $\pmb q^{(n)}$ of convex problem (\ref{alg2min2}).
\STATE With fixed $( \pmb q^{(n)}, \pmb d^{(n-1)})$, obtain the optimal $\pmb t^{(n)}$ of convex problem (\ref{alg2min2}).
\STATE With fixed $( \pmb q^{(n)}, \pmb t^{(n)})$, obtain the optimal $\pmb d^{(n)}$ of convex problem (\ref{alg2min2}).
\STATE Set $n=n+1$.
\UNTIL the objective function (\ref{alg2min2}a) converges.
\end{algorithmic}
\end{algorithm}

According to Algorithm 1,
the complexity of the proposed algorithm lies in solving three convex problems.
Since the dimension of variable $\pmb q$ is $N$, the complexity of solving problem (\ref{alg2min2}) with fixed ($\pmb t, \pmb d$) by using the standard interior point method
\cite[Pages 487, 569]{boyd2004convex} is $\mathcal O(N^3)$.
With the same analysis, the complexities of solving $\pmb t$ and $\pmb d$ are $\mathcal O(N^3)$ and  $\mathcal O(M^3)$, respectively.
Since the number of groups is less than the number of users, i.e., $N<M$, the total complexity for solving problem (\ref{alg2min2}) is $\mathcal O(LM^3)$, where $L$ denotes the total number of iterations of Algorithm 1.

\section{Numerical Results}

In this section, numerical results are presented  to evaluate the performance of the proposed algorithm.
The MEC system consists of $M=20$ users.
The path loss model is $128.1+37.6\log_{10} d$ ($d$ is in km)
and the standard deviation of shadow fading is $4$ dB \cite{access2010further,olfat2008optimum,
shadmand2010cross,shojaeifard2011joint,7947159,7264975}.
In addition, the bandwidth of the system is $B=10$ MHz, and the noise power is $\sigma^2=-104$ dBm.
For MEC parameters, the data size and the required number of CPU cycles per bit are  set to follow equal distributions with $R_{j} \in[100,500]$ Kbits and $C_{j}\in[500, 1500]$ cycles/bit.
The CPU computation of each user is set as the same $F_{j}=1$ GHz and the local computation energy per cycle for each user or the BS is also set as equal $P_{j}=10^{-10}$ J/cycles for all $j \in \mathcal M$ and $j=0$.
The self-interference coefficient at the BS is $\gamma=10^{-5}$ and the energy efficiency of the EH process for each user is $\zeta_j=0.8$ .
Besides, the maximal transmission power of each user and the BS are respectively set as $P=30$ dBm and $Q=47$ dBm.
Unless specified otherwise, the system parameters are set as time solt duration $T=0.1 $ s, and the edge computation capacity $F=6\times 10^9$ cycles per slot.

\begin{figure}
\centering
\includegraphics[width=3.0in]{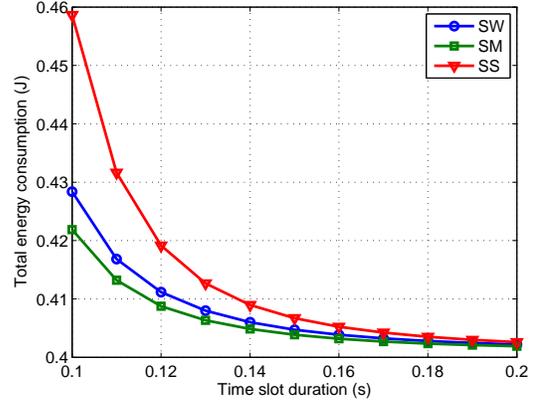}
\vspace{-1em}
\caption{Impact of user pairing on the total energy consumption of the proposed algorithm.}\label{fig5}
\vspace{-0.5em}
\end{figure}

Due to decoding complexity and error propagation, it is  recommended that each resource is multiplexed by small number of users (for example, two users) \cite{6464495}.
In simulations, we consider that each group has two users.
We study the influence of user pairing by considering three different user-pairing methods \cite{7273963}.
For strong-weak (SW) pair selection, the user with the strongest channel condition is paired with the user with the weakest, and
the user with the second strongest is paired with one with the second weakest, and so on.
For strong-middle (SM) pair selection, the user with the strongest channel condition is paired with the user with the middle strongest user, and so on.
For strong-strong (SS) pair selection, the user with the strongest channel condition is paired with the one with the second strongest, and so on.
In Fig. \ref{fig5}, we show the total energy consumption of the proposed algorithm.
It is observed that SM outperforms the other two methods in terms of total energy consumption.
This is due to the fact that two users in any group of the SM scheme have relative large channel gain difference.
Due to the superiority of SM, the following simulations are based on SM pair selection.

\begin{figure}
\centering
\includegraphics[width=3.0in]{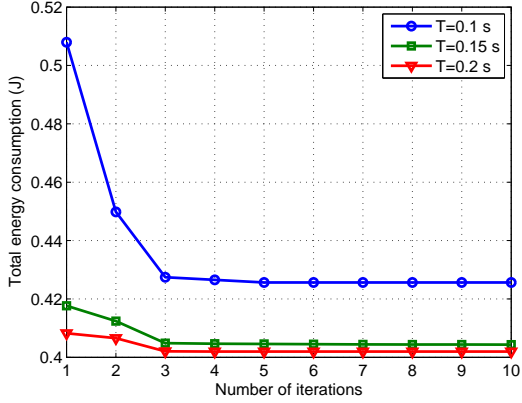}
\vspace{-1em}
\caption{Convergence behaviour of the proposed algorithm under different cloud computation capacities.}\label{fig1}
\vspace{-0.5em}
\end{figure}

Fig. \ref{fig1} illustrates the convergence behaviours for the proposed algorithm under different cloud computation capacities.
It can be seen that the proposed algorithm converges rapidly, and only three times are sufficient to converge, which shows the effectiveness of the proposed algorithm.

\begin{figure}
\centering
\includegraphics[width=3.0in]{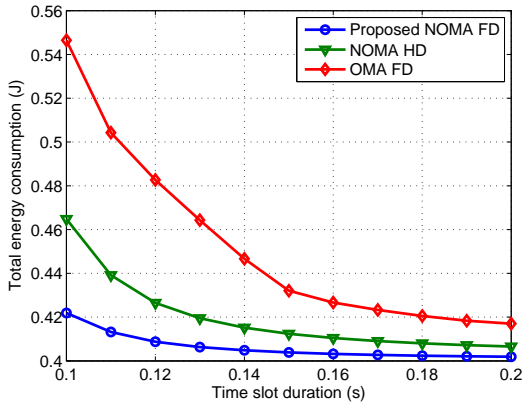}
\vspace{-1em}
\caption{Total energy consumption versus time solt duration.}\label{fig2}
\vspace{-0.5em}
\end{figure}

We compare the total energy consumption performance of the proposed algorithm (labelled as `Proposed NOMA FD') with the algorithm for a half-duplex (HD) MEC system with NOMA \cite{8269088} (labelled as `NOMA HD'), and the algorithm for a FD-MEC system with OMA \cite{8254208} (labelled as `OMA FD').

The total energy consumption versus time slot duration is depicted in Fig.~\ref{fig2}.
From this figure, we find that the total energy consumption decreases with time slot duration.
It can be shown that the proposed algorithm yields best performance among all algorithms.
Since FD enables simultaneously energy transfer and data reception, the proposed algorithm yields lower energy consumption than NOMA HD.
Compared with OMA FD, NOMA reduces the total energy consumption of all users at the cost of adding computing complexity at the BS due to SIC.

\begin{figure}
\centering
\includegraphics[width=3.0in]{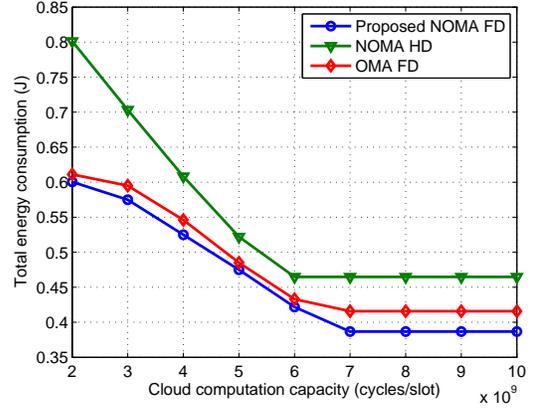}
\vspace{-1em}
\caption{Total energy consumption versus cloud computation capacity.}\label{fig3}
\vspace{-0.5em}
\end{figure}

In Fig.~\ref{fig3}, we show the total energy consumption versus cloud computation capacity.
It is observed that the total energy consumption decreases with cloud computation capacity since higher cloud computation capacity allows users to offload more data to the BS, resulting lower energy consumption at users.
The proposed algorithm achieves the best performance according to this figure, which shows the effectiveness of the proposed algorithm.
Besides, the total energy consumption keeps stable when the cloud computation capacity exceeds a threshold which coincides with previous findings in \cite{7842016}.

\section{Conclusion}

In this paper, we have investigated the total energy minimization problem for a FD-MEC system with NOMA and EH.
The original nonconvex problem is first equivalent to a problem with less variables.
Then, an iterative algorithm is accordingly proposed with low complexity.
Numerical results show that it is energy efficient to pair the user with the strongest channel condition is paired with the user with the middle strongest user and the proposed algorithm achieves better performance than conventional schemes in terms of energy consumption.


\appendices
\section{Proof of Theorem 1}
\setcounter{equation}{0}
\renewcommand{\theequation}{\thesection.\arabic{equation}}
According to Lemma 1, setting offloading constraints (\ref{sysmin1}b) with equality yields
\begin{equation}\label{appenAeq1}
2^{\frac{d_{j}}{Bt_i}} \sum_{l=j+1}^{J_i} h_{l}p_l
+
(\sigma^2+\gamma q_i)\left(2^{\frac{d_{j}}{Bt_i}} -1\right)
=  \sum_{l=j}^{J_i} h_{l} p_l,
\end{equation}
for $j=J_{i-1}+1, \cdots, J_i$.
To solve equations (\ref{appenAeq1}), we define
\begin{equation}\label{appenAeq2}
u_{j}=\sum_{l=j}^{J_i} h_{l} p_{l}, \quad \forall j\in\mathcal  J_i,
\end{equation}
which is expressed as the summation of transmission power multiplied by from user $j$ in group $i$ to the last user $J_i$ in group $i$.
Based on (\ref{appenAeq1}) and (\ref{appenAeq2}), we can obtain
\begin{equation}\label{appenAeq2_1}
{u_{j}}={2^{\frac{d_{j}}{Bt_i}}} u_{{j+1}}+
(\sigma^2+\gamma q_i)
 \left(2^{\frac{d_{j}}{Bt_i}} -1\right), \quad \forall j\in \mathcal J_i .
\end{equation}
Due to the fact that $\sum_{l=J_i+1}^{J_i}p_{j}=0$, we have
\begin{equation}\label{appenAeq2_2}
u_{J_i+1}=0.
\end{equation}
Based on (\ref{appenAeq2_2}), we solve (\ref{appenAeq2_1}) by using the recursion method and obtain
\begin{equation}\label{appenAeq3}
u_j=(\sigma^2+\gamma q_i)\sum_{l=j}^{J_i}
\left(2^{\frac{d_{l}}{Bt_i}} -1\right) {2^{\frac{\sum_{s=j}^{l-1}d_{s}}{Bt_i}}}, \quad \forall  j\in \mathcal J_i,
\end{equation}
where we define $ {2^{\sum_{s=j}^{j-1}\frac{d_{s}}{B t_i}}}=2^0$.

From (\ref{appenAeq2}) and (\ref{appenAeq2_2}), we can obtain the transmission power of user $j$ as
\begin{eqnarray}\label{appenAeq3_2}
 \!\!\!p_{j} &&\!\!\!\!\!\!\!\!\!\!
 =\frac{u_{j}-u_{j+1}}
{h_{j}}
\nonumber \\
 &&\!\!\!\!\!\!\!\!\!\!
=\sum_{l=j}^{J_i}
\frac{(\sigma^2+\gamma q_i)}
{h_{j} }
 \left(2^{\frac{d_{l}}{B t_i}} -1\right){2^{\frac{\sum_{s=j}^{l-1}d_{s}}{B t_i}}}
\nonumber \\
 &&\!\!\!\!\!\!\!\!\!\!\quad-
\sum_{l=j+1}^{J_i}
\frac{(\sigma^2+\gamma q_i)}
{h_{j}}
\left(2^{\frac{d_{l}}{B t_i}} -1\right) {2^{\frac{\sum_{s=j+1}^{l-1}d_{s}}{B t_i}}}
\nonumber\\
&&\!\!\!\!\!\!\!\!\!\!
=f_{ij}(t_i,q_i,\pmb d).
\end{eqnarray}
Substituting (\ref{appenAeq3_2}) into problem (\ref{sysmin1}) yields the equivalent problem (\ref{alg2min2}).

\section{Proof of Theorem 2}
\setcounter{equation}{0}
\renewcommand{\theequation}{\thesection.\arabic{equation}}

We first prove that problem (\ref{alg2min2}) is convex in $\pmb q$ with fixed $(\pmb t, \pmb d)$.
Based on (\ref{alg2min2}), $f_{ij}(t_i,q_i,\pmb d)$ is a linear function of $q_i$ with fixed $(\pmb t, \pmb d)$.
Because the objective function and all constraints of problem (\ref{alg2min2}) are linear with fixed $(\pmb t, \pmb d)$, problem (\ref{alg2min2}) is a linear problem (also convex problem) with fixed $(\pmb t, \pmb d)$.

We then prove that problem (\ref{alg2min2}) is convex in $\pmb t$ with fixed $(\pmb q, \pmb d)$.
To show this, we define a function
\begin{eqnarray}\label{appenBeq1}
u (x)= { ( {\text e}^{ {a x} } - 1 )
 ( {\text e^{ {b x} }} -  1 )
\text e^{  {c x} }}, \quad \forall x\geq 0.
\end{eqnarray}
Then, the second-order derivative follows
\begin{eqnarray}\label{appenBeq2}
u''(x)= &&\!\!\!\!\!\!\!\!\!\!
(a^2+2 ac ) ( {\text e^{ {bx} }} - 1 )\text e^{ ({a +c})x}
\nonumber\\
&&\!\!\!\!\!\!\!\!\!\!
+2ab  \text e^{ ({a +b+ c })x}
\nonumber\\
&&\!\!\!\!\!\!\!\!\!\!
+( b^2+2bc)  ( {\text a^{ {a x} }} - 1 )\text e^{ ({b+c})x}
\nonumber\\
&&\!\!\!\!\!\!\!\!\!\!
+c^2
( {\text e}^{ {a x} } - 1 )
( {\text e^{ {b x} }} - 1 )\text e^{  c x}\geq 0,
\end{eqnarray}
which shows that $u (x)$ is a convex function in $x$.
According to \cite[Page~89]{boyd2004convex}, the perspective of $u(x)$ is the function $v(x,t)$ defined by $v(x,t)=tu(x/t)$, ${\textbf{dom}}\:v= \{(x,t)|x/t\in {\textbf{dom}}\:u, t>0\}$.
If $u(x)$ is a convex function, then so is its perspective function $v(x,t)$ \cite[Page~89]{boyd2004convex}.
Then, $v(x,t)=tu(x/t)$ is convex in $(x,t)$, and $v(1,t)$ is also convex in $x$.
As a result, $t_i f_{ij}(t_i, q_i, \pmb d)$ is convex in $t_i$.
Due to the fact that $f_{ij}(t_i, q_i, \pmb d)$ is a decreasing function of $t_i$, $f_{ij}(t_i, q_i, \pmb d) \leq P_i$ can be equivalent to a linear equation $t_i \geq t_{i}^{\min}$, where $f_{ij}(t_i^{\min}, q_i, \pmb d) = P_i$.
Because the objection function and all the constraints of problem (\ref{alg2min2}) are convex, problem (\ref{alg2min2}) is convex in $\pmb t$ with fixed $(\pmb q, \pmb d)$.

Finally, we show that problem (\ref{alg2min2}) is convex in $\pmb d$ with fixed $(\pmb q, \pmb t)$.
From (\ref{alg2eq2}), $f_{ij}(t_i, q_i ,\pmb d) $ is convex in $d_i$.
Based on this finding, we can prove that problem (\ref{alg2min2}) is convex in $\pmb d$ with fixed $(\pmb q, \pmb t)$.

\bibliographystyle{IEEEtran}
\bibliography{IEEEabrv,MMM}
\end{document}